\documentclass{jps-cp}
\usepackage{txfonts} %Please comment out this line unless the txfonts package is availabe in your LaTeX system.

\usepackage{color}

\def\ie{{\sl i.e.}}

\def\ptv{p_{\rm{T}}}

\def\glgt{|G_{\rm L}/G_{\rm T}|}
\def\thetalt{\theta_{\rm LT}}
\def\Mhh{M_{12}}

\def\gl{G_{\rm L}}
\def\gt{G_{\rm T}}

\title{Simulation of the Polarized Quark Fragmentation with Vector Meson Production}

\author{Albi \textsc{Kerbizi}$^{1,*}$, Xavier \textsc{Artru}$^{2}$ and Anna \textsc{Martin}$^{1}$}

\inst{$^{1}$ INFN Sezione di Trieste and Dipartimento di Fisica, Universit\`a degli Studi di Trieste, \\
Via Valerio 2, 34127 Trieste, Italy\\
$^{2}$ Universit\'e de Lyon, Institut de Physique des deux Infinis (IP2I Lyon), Universit\'e Lyon 1 and CNRS, 4 rue Enrico Fermi, F-69622 Villeurbanne, France\\
$^{*}$ Speaker.}

\email{albi.kerbizi@ts.infn.it}

\recdate{January 15, 2022}

\abst{The recently developed quantum mechanical string+${}^3P_0$ model of polarized quark fragmentation with pseudoscalar and vector meson production has been implemented in a stand alone Monte Carlo program. The simulation enables the detailed study of the quark spin dependent effects in the fragmentation process such as the transverse spin asymmetries. In particular, it allows to investigate the effects of the production and decay of vector mesons on the Collins and dihadron asymmetries of the observed hadrons, and to study the still unknown Collins effect for vector mesons. The main results of the Monte Carlo simulations obtained so far are summarized and the comparison with the existing experimental data on SIDIS and $e^+e^-$ annihilation to hadrons processes are presented.}

\kword{quark, fragmentation, vector mesons, polarization, Collins effect, string+3P0}

\begin{document}
\maketitle

\section{Introduction}
The dependence on the quark spin of the fragmentation process is an interesting subject in its own right and essential to access the transverse momentum and transverse spin structure of the nucleons. Transverse spin-dependent effects such as the Collins effect \cite{Collins-FF} and the dihadron asymmetry \cite{Bianconi-2h} behave as polarimeters of the quark transverse polarization in the nucleon, and have been used in semi inclusive deep inelastic scattering (SIDIS) process to access the quark transversity parton distribution function (PDF). They can also be utilized to analyze other still unknown leading twist transverse momentum dependent (TMD) asymmetries. To extract information on TMD PDFs from SIDIS data, however, the Collins and dihadron analyzing powers must be taken from the corresponding asymmetries in $e^+e^-$ annihilation to hadrons process.

The spin-dependent fragmentation process can be tackled also through models and their implementation in Monte Carlo (MC) event generators which enable the calculation of the analysing powers.
A solid model of the fragmentation process of polarized quarks is the string+${}^3P_0$ model which extends the Lund string fragmentation model \cite{Lund-model} by the addition of the quark spin degree of freedom. Two slightly different versions of the model restricted to pseudoscalar mesons, M18 in \cite{Kerbizi-2018} and M19 in \cite{Kerbizi-2019}, have been proposed and have been implemented in stand alone MC programs which give similar results. The simpler version M19 has been interfaced to the Pythia 8 hadronization \cite{Pythia-8} via the StringSpinner \cite{StringSpinner} package which allows the simulation of the Collins and dihadron asymmetries in SIDIS.

The present article focuses on a recently developed version of the string+${}^3P_0$ model (M20) which extends M19 by systematically introducing vector mesons. The theoretical aspects of the string+${}^3P_0$ model and of M20 are subject of a different article in these proceedings \cite{these-proceedings}. A detailed description of the model and of the simulation results can be found in Ref. \cite{Kerbizi-2021}. Here the main results obtained from stand alone MC simulations with M20 are presented.

A brief introduction to the structure of the MC implementation of M20 is given in Sec. \ref{sec:MC program}. In Sec. \ref{sec:TSA} the effects of the introduction of vector mesons in the polarized quark fragmetation chain on the Collins and dihadron asymmetries as well as the Collins asymmetry for the $\rho$ mesons are shown. The comparison with SIDIS and $e^+e^-$ data is given in Sec. \ref{sec:comparison}. Finally the conclusions are drawn.

\section{Structure of the Monte Carlo program}\label{sec:MC program}
The stand alone MC program simulates the fragmentation event $q_A\,\bar{q}_B\rightarrow h_1,\dots,h_r,\dots, h_N$ of a string stretched between the quark $q_A$ and the remnant $\bar{q}_B$ (an anti-quark in an $e^+e^-$ annihilation or the target remnant in SIDIS), in the produced hadrons $h_r$ with $r=1,2,\dots,N$. The index $r$ indicates the rank of the hadron ($h_1$ contains $q_A$), which can be either a pseudoscalar meson (PSM = $\pi,K,\eta,\eta'$) or a vector meson (VM = $\rho,K^*,\omega,\phi$).

After defining the flavor, the momentum and the $2\times 2$ spin density matrix $\rho(q_A)$ of $q_A$, the fragmentation chain is simulated by recursively repeating the elementary splitting $q_r\rightarrow h_r+q_{r+1}$ of a quark $q_r$ in the hadron $h_r$ and the leftover quark $q_{r+1}$. The splitting consists in the following steps: (i) generate a $q_{r+1}\bar{q}_{r+1}$ pair, (ii) form the hadron $h_r(q_r\bar{q}_{r+1})$, (iii) assign $h_r$ to the PSM or to the VM multiplet, (iv) generate the hadron four momentum, (iv) for a vector $h_r$, calculate the spin density matrix, simulate the decay and calculate the decay matrix (following the recipe in Refs. \cite{Collins-corr,Knowles-corr} to take into account the spin correlations between $h_r$ and $q_{r+1}$),(vi) calculate the spin density matrix of $q_{r+1}$. The steps (i)-(vi) are iterated until the squared transverse mass of the remaining system $q_{N+1}\bar{q}_B$ falls below $1.5\,(GeV/\rm{c}^2)^2$. Then the fragmentation chain is terminated.

The simulation of steps (i)-(vi) is based on the splitting amplitude $T_{q_{r+1},h_r,q_r}$, which governs the splitting in flavor, momentum and spin spaces. It combines the Lund string fragmentation model, the quark $2\times2$ spin matrices arising from the assumption that the $q_{r+1}\bar{q}_{r+1}$ pair at string breaking are produced in the ${}^3P_0$ state, and the coupling of quarks to the produced meson. The ${}^3P_0$ mechanism is parametrized by a complex mass parameter $\mu$ taken as in Ref. \cite{Kerbizi-2019}, the imaginary part of which is responsible for the transverse spin effects.
The coupling of quarks to VMs with linear and transverse polarization with respect to the string axis (defined by the momentum of $q_A$ in the $q_A\bar{q}_B$ rest frame), is parametrized respectively by the complex parameters $\gl$ and $\gt$. The actual free parameters are $2|\gt|^2+|\gl|^2$, which gives the relative fraction of VM respect to PSM, $\glgt$, which governs the Collins effect for VMs, and $\thetalt=\arg{\gl/\gt}$ which is responsible for the oblique polarization of VMs and acts as a new source of Collins effect on the decay products of the VM.

%In the following sections we consider results obtained from simulations of fully transversely polarized $u$ quarks. More details can be found in Ref. \cite{Kerbizi-2021}.

\section{Results for transverse spin asymmetries}\label{sec:TSA}

%\begin{figure}[tbh]
%\centering
%\begin{minipage}[b]{0.48\textwidth}
%\includegraphics[width=1.0\textwidth]{CollinsM19M20_4.eps}
%\end{minipage}
%\begin{minipage}[b]{0.48\textwidth}
%\includegraphics[width=1.0\textwidth]{CollinsM19M20_4.eps}
%\end{minipage}
%\caption{ aa}
%\label{f1}
%\end{figure}

\subsection{Effect of vector mesons on the Collins and dihadron asymmetries}
\begin{figure}[tbh]
\centering
\begin{minipage}[b]{0.48\textwidth}
\includegraphics[width=1.1\textwidth]{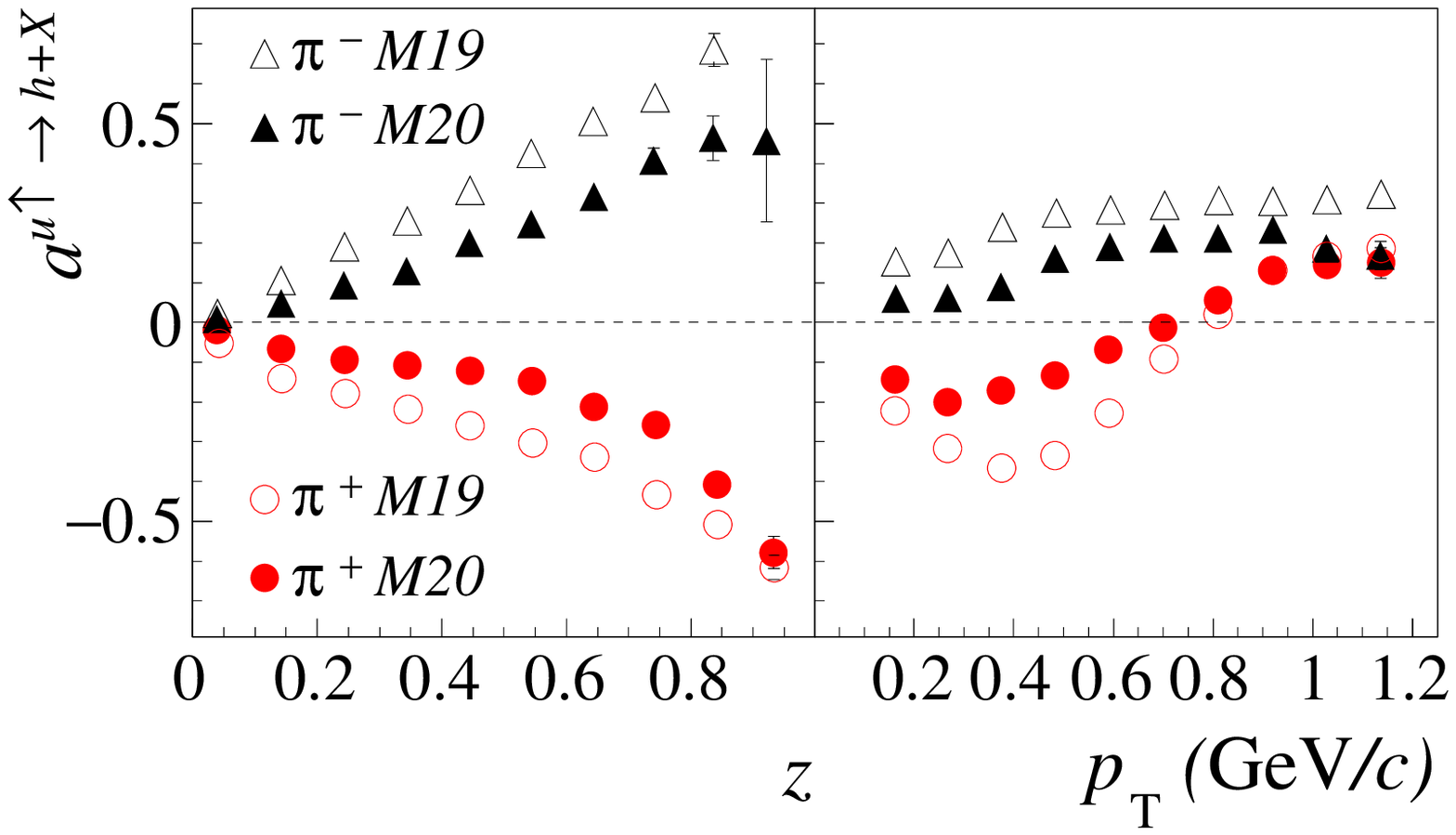}
\end{minipage}
\begin{minipage}[b]{0.48\textwidth}
\includegraphics[width=1.1\textwidth]{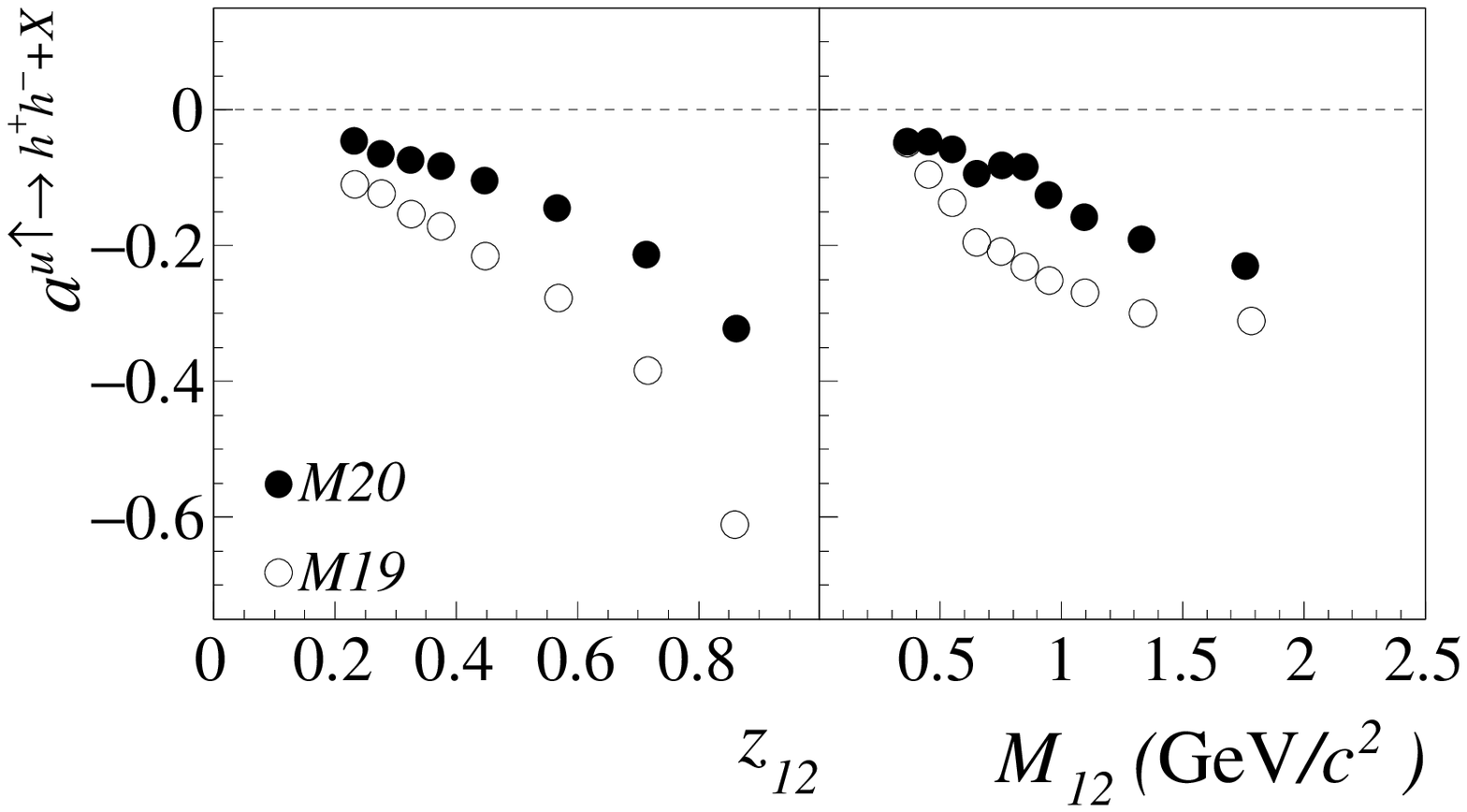}
\end{minipage}%\vspace{-1em}
\caption{Left figure: Collins analysing power for $\pi^+$ (circles) and $\pi^-$ (triangles) as function of $z$ and $\ptv$. Right figure: dihadron analysing power for $h^+h^-$ pairs as function of $z_{12}$ and $\Mhh$. The closed points show the results obtained with M20 whereas the open points show those obtained with M19.}
\label{fig:Effect on Collins and 2h}
\end{figure}

The Collins analysing power for positive and negative pions as obtained with M20 in fragmentations of fully transversely polarized $u$ quarks with $\glgt=1$ and $\thetalt=0$ is shown by the closed points in the left plot of Fig. \ref{fig:Effect on Collins and 2h}. It is calculated as $a^{u\uparrow\rightarrow h+X}=2\langle \sin(\phi_h-\phi_{S_q})\rangle$ as function of the hadron fractional energy $z$ and of the transverse momentum $\ptv$ with respect to the fragmenting quark momentum. The cuts $z>0.2$ (when looking at the $\ptv$ dependence) and $\ptv>0.1 \rm(GeV/c)$ have been applied. The angles $\phi_h$ and $\phi_{S_q}$ are the azimuthal angles of the hadron transverse momentum and of the fragmenting quark transverse polarisation respectively. In the same plot, the Collins analysing power as obtained with M19 shown for comparison. As can be seen the effect of the production and decay of vector mesons in the Collins analysing power of the observed hadrons is large as function of both kinematic variables in the whole kinematic range. The average analysing power as obtained with only pseudoscalar mesons is reduced by a factor of two.

A similar effect can be seen also on the dihadron analysing power for oppositely charged hadron pairs in the same fragmentation event, shown in the right plots of Fig. \ref{fig:Effect on Collins and 2h}. The dihadron analysing power is calculated as $a^{u\uparrow\rightarrow h^+h^- +X}=2\langle \sin_(\phi_R-\phi_{S_q})\rangle$ as function of the fractional energy of the pair $z_{12}$ and of their invariant mass $M_{12}$. $\phi_R$ is the azimuthal angle of the relative momentum of the hadron pair $\textbf{R}$. Its transverse component with respect to fragmenting quark momentum is given by $\textbf{R}_T=z_2\textbf{p}_{1\rm T}/z_{12} - z_1\textbf{p}_{2\rm T}/z_{12}$, the indices $1$ and $2$ referring respectively to the positively and negatively charged hadron of the pair. The kinematic cuts $z_{1,2}>0.1$, $R_{\rm T}>0.07\,\rm{GeV}/c$ and $x_{F\,1,2}>0.1$ have been applied, with $x_F$ being the Feynman variable. The decays of the VMs are invariant under $\textbf{R} \leftrightarrow -\textbf{R}$ and do not contribute to the dihadron analysing power, which is diluted as compared to M19. The effect is particularly clear as function of $\Mhh$ around the $\rho^0(770)$ meson mass.

If the hadrons of the pair are ordered according to their fractional energies (\ie{} $z_1>z_2$) a new dihadron asymmetry shows up Ref.\cite{Kerbizi-2021}. The measurement of such "$z$-ordered asymmetry" would be interesting to shed more light on the mechanism of VM production in polarized quark fragmentations.

\subsection{Collins asymmetry for $\rho$ mesons}
\begin{figure}[tbh]
\centering
%\vspace{-4em}
\includegraphics[width=0.45\textwidth]{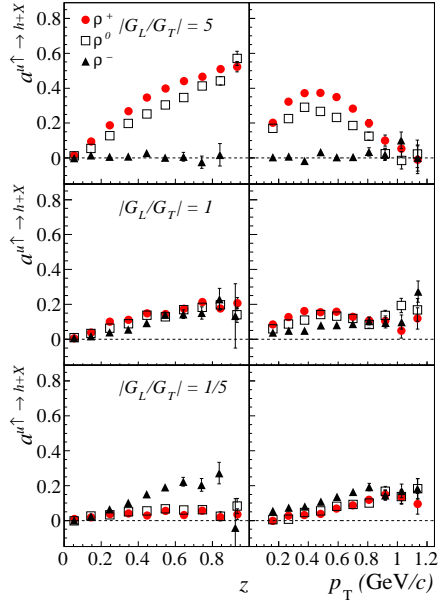}
\caption{Collins asymmetry for $\rho^+$ (circles), $\rho^0$ (squares) and $\rho^-$ (triangles) as function of $z$ and $\ptv$. The top panel is obtained with $\glgt=5$, the middle panel with $\glgt=1$ and the lower panel with $\glgt=1/5$. In all cases the value $\thetalt=0$ has been used.}
\label{fig:rho mesons}%\vspace{-2em}
\end{figure}

The Collins analyzing power for $\rho$ mesons is shown in Fig. \ref{fig:rho mesons} as function of $z$ and $\ptv$ for different values of the free parameter $\glgt=5,1,1/5$. These values favor respectively the coupling of quarks to longitudinally polarized, unpolarized and transversely polarized VMs. The parameter $\thetalt$ has been taken to be vanishing since it does not have effect on the Collins analysing power for VMs.

For all values of $\glgt$ the analysing power of $\rho$ mesons has opposite analysing power as compared to the positive pions shown in the left panel in Fig. \ref{fig:Effect on Collins and 2h}, in agreement with the prediction in Ref. \cite{Czyzewski}. Large values of the analysing power for $\rho^+$ and $\rho^0$, comparable with that of the pions in Fig. \ref{fig:Effect on Collins and 2h},  are obtained for $\glgt=5$, namely for longitudinally polarized vector mesons. For the same value of $\glgt$ the analysing power of $\rho^-$ mesons vanishes. This hierarchy is reversed for $\glgt=1/5$ for which the analysing power of $\rho^+$ and $\rho^0$ is smaller than that of the $\rho^-$. For $\glgt=1$ the $\rho$ mesons have nearly the same analysing power.

The first measurement of the Collins asymmetry for inclusive $\rho^0$ mesons produced in SIDIS has been performed by COMPASS \cite{rho0-TSA} demonstrating that the measurement of this asymmetry is feasible. Despite the large uncertainties, a first comparison of the simulated Collins asymmetry for $\rho^0$ mesons with the COMPASS results seems to favor $\glgt=5$, hence the coupling of quarks with longitudinally polarized VMs.

\section{Comparison with SIDIS and $e^+e^-$ data}\label{sec:comparison}
\begin{figure}[tbh]
\vspace{-4em}
\centering
\begin{minipage}[b]{0.48\textwidth}
\includegraphics[width=1.0\textwidth]{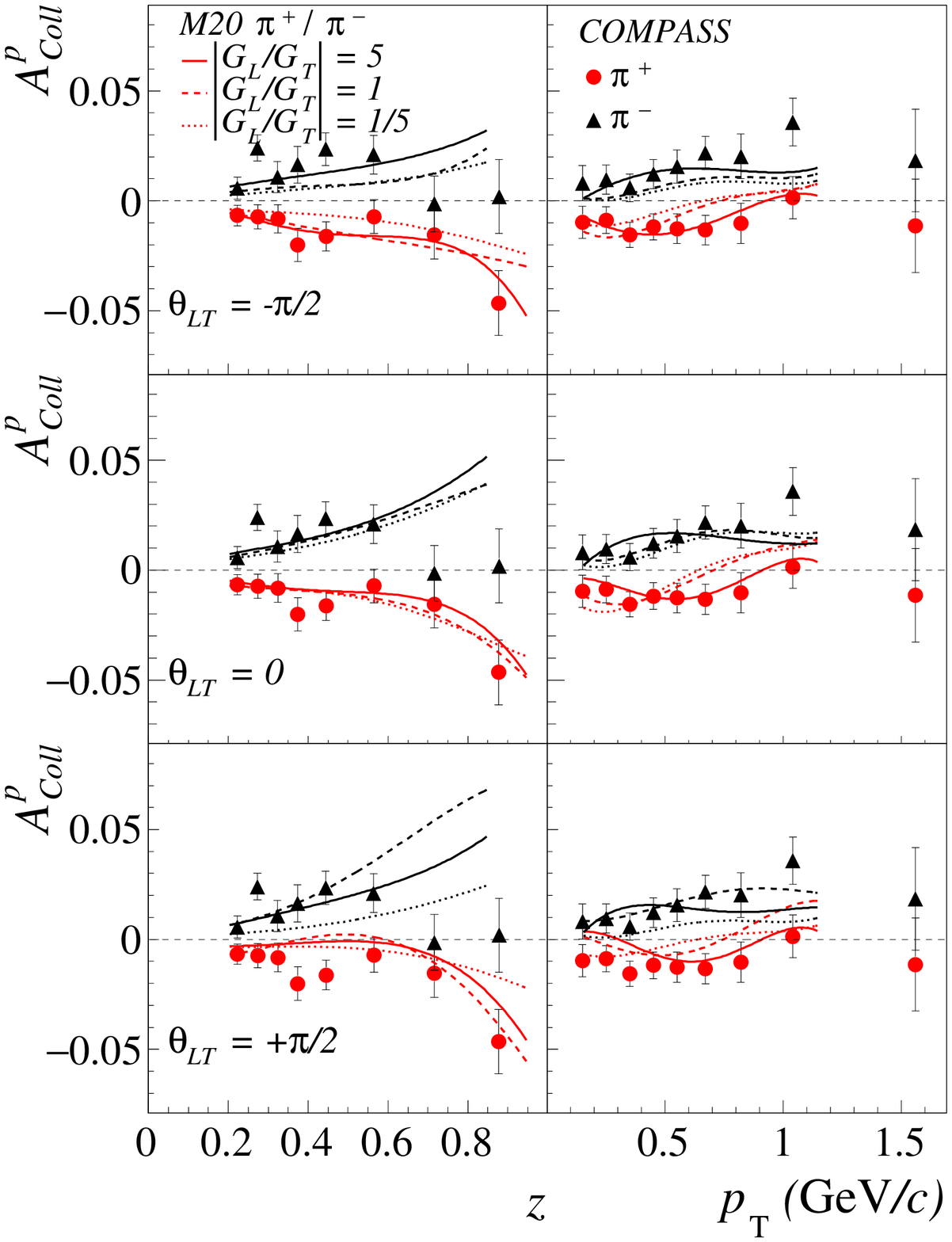}
\end{minipage}
\begin{minipage}[b]{0.48\textwidth}
\includegraphics[width=1.0\textwidth]{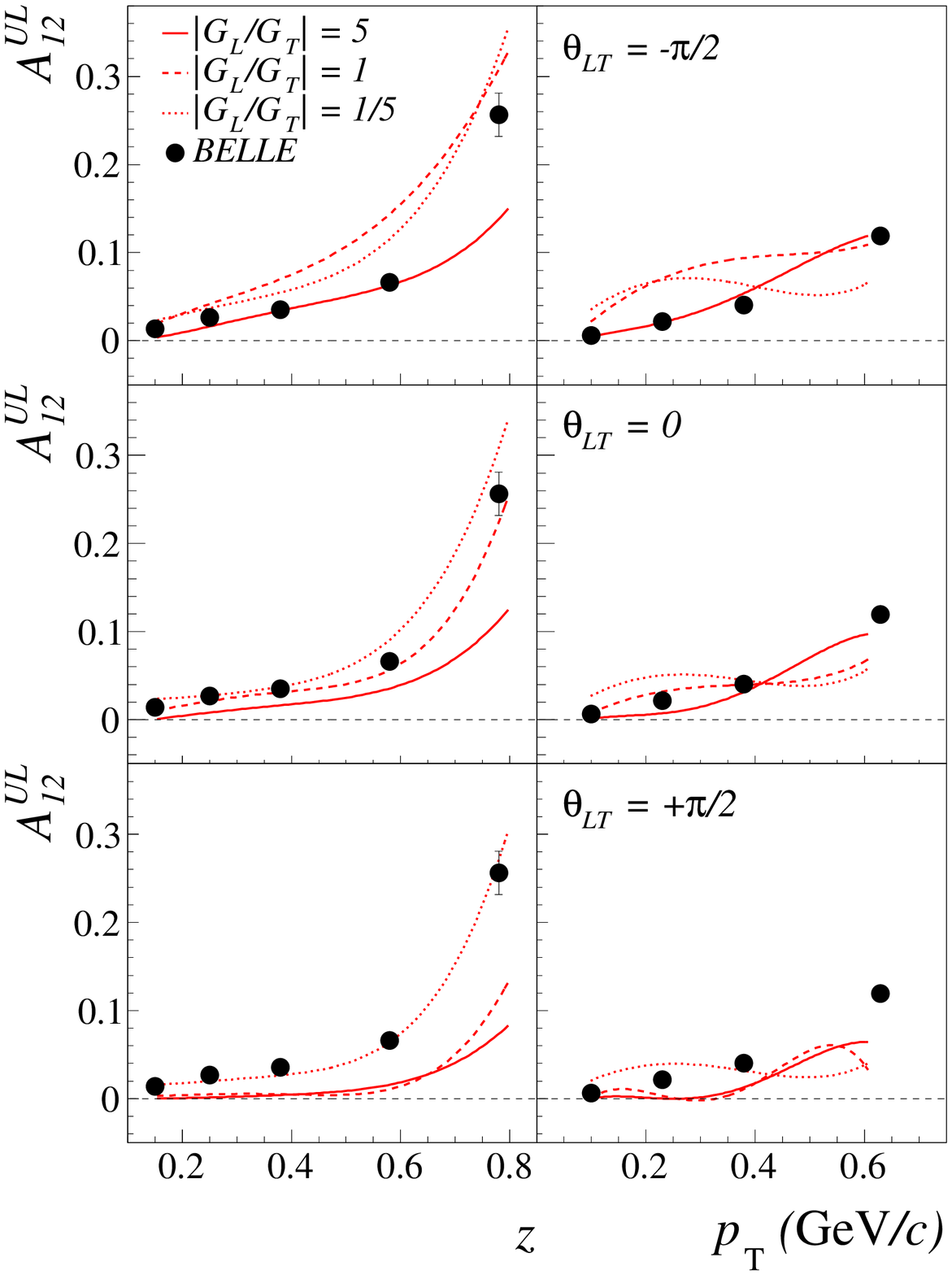}
\end{minipage}
\caption{Comparison between the simulation results for different values of $\glgt$ and $\thetalt$ (lines) with the experimental data (points) on the Collins asymmetries for charged pions measured by COMPASS in SIDIS off transversely polarized protons \cite{Compass-Collins} (left figure) and the Collins asymmetries for back-to-back oppositely charged pions measured by BELLE in $e^+e^-$ annihilation to hadrons \cite{Belle-Collins} (right figure).}
\label{fig:comparison_with_data}
\end{figure}

To constrain the values of the free parameters $\glgt$ and $\thetalt$ we have compared the simulation results with the Collins asymmetries for $\pi^+$ and $\pi^-$ measured by COMPASS in SIDIS off a transversely polarized proton target \cite{Compass-Collins}. The simulated Collins asymmetries are shown by lines in the left plot of Fig. \ref{fig:comparison_with_data} as function of $z$ and $\ptv$. The top panel shows the results for $\thetalt=-\pi/2$, the middle panel for $\thetalt=0$ and the lower panel for $\thetalt=+\pi/2$. In each panel the asymmetries obtained for $\glgt=5$, $1$ and $1/5$ are shown. For each combination of the free parameter values, the simulated analysing power has been scaled by a constant factor to take into account the partial polarization of quarks described by the transversity PDF. As can be seen the effect of oblique polarization given by non-zero $\thetalt$ is large particularly as function of $z$. A satisfactory description of the data is generally obtained for $\glgt\geq 1$ and $\thetalt \leq 0$. A precise determination of the parameter values is however not possible, given the large uncertainties of the data. A chi-square test suggests the combinations $\glgt=1$ and $\thetalt=0$, or $\glgt=5$ and $\thetalt=-\pi/2$.

Similar conclusions can be drawn when looking at the right plot of Fig. \ref{fig:comparison_with_data}, which shows the comparison between the simulation results with the Collins asymmetries for oppositely charged back-to-back pion pairs produced in $e^+e^-$ annihilation events at BELLE \cite{Belle-Collins}. The asymmetries are shown as function of the fractional energy $z$ and of the transverse momentum $\ptv$ with respect to the string axis, approximated in the data by the thrust axis. Following Ref. \cite{Belle-Collins}, the data has been corrected for the charm quark contamination factors assuming vanishing asymmetries in charm initiated events.
Results obtained in simulations with values $\glgt=5$ and $\thetalt=-\pi/2$ provide a satisfactory comparison with the $e^+e^-$ data. Like in the SIDIS case, however, the combination $\glgt=1$ and $\thetalt=0$ gives also a satisfactory description of data and can not be excluded.
\section{Conclusions}
The string+${}^3P_0$ model of polarized quark fragmentation has been recently extended by systematically introducing the production of vector mesons \cite{Kerbizi-2021}. The new model has few free parameters and has been implemented in a stand alone Monte Carlo program. Combinations of possible values of the free parameters $\glgt$ and $\thetalt$ have been selected by comparing the simulation results with the SIDIS and $e^+e^-$ transverse spin asymmetry data, and a satisfactory description has been found. Also, the simulations produce large values of the Collins analysing power for $\rho$ mesons. More precise measurements of the Collins asymmetries for vector mesons in SIDIS and their measurement in $e^+e^-$ annihilation to hadrons would help in determining the values of the free parameters of the model.

\end{document}